\documentclass[a4paper,fleqn,usenatbib]{mnras}

\usepackage{newtxtext,newtxmath}
\usepackage[T1]{fontenc}
\usepackage{ae,aecompl}

\usepackage{graphicx}	% Including figure files
\usepackage{amsmath}	% Advanced maths commands
 \usepackage{amssymb}	% Extra maths symbols
\usepackage{color}

\title{The last migration trap of compact objects in AGN accretion disc}

\author[P. Peng, X. Chen]{Peng Peng$^{1}$, Xian Chen$^{1,2}\thanks{E-mail: \href{mail to: xian.chen@pku.edu.cn}{xian.chen@pku.edu.cn}} $
\\
$^{1}$Astronomy Department, School of Physics, Peking University, 100871 Beijing, China
\\
$^{2}$Kavli Institute for Astronomy and Astrophysics at Peking University, 100871 Beijing, China
}

\date{Accepted XXX. Received YYY; in original form ZZZ}

\pubyear{2020}

\begin{document}
\label{firstpage}
\pagerange{\pageref{firstpage}--\pageref{lastpage}}
\maketitle

% Abstract of the paper
\begin{abstract}
Many black holes (BHs) detected by the Laser Interferometer Gravitational-wave
Observatory (LIGO) and the Virgo detectors are multiple times more massive than
those in X-ray binaries.  One possibility is that some BBHs merge within a
few Schwarzschild radii of a supermassive black hole (SMBH), such that the
gravitational waves (GWs) are highly redshifted, causing the mass inferred from
GW signals to appear higher than the real mass. The difficulty of this scenario
lies in the delivery of BBH to such a small distance to a SMBH.  Here we
revisit the theoretical models for the migration of compact objects (COs) in
the accretion discs of active galactic nuclei (AGNs).  We find that when the
accretion rate is high so that the disc is best described by the slim disc
model, the COs in the disc could migrate to a radius close to the innermost
stable circular orbit (ISCO) and be trapped there for the remaining lifetime
of the AGN.  The exact trapping radius coincides with the transition region
between the sub- and super-Keplerian rotation of the slim disc. We call this region
``the last migration trap'' because inside it COs
can no longer be trapped for a long time.  
We pinpoint the parameter space which could induce
such a trap and we estimate that the last migration trap contributes a few per
cent of the LIGO/Virgo events. Our result implies that a couple of BBHs
discovered by LIGO/Virgo could have smaller intrinsic masses.
\end{abstract}

% Select between one and six entries from the list of approved keywords.
% Don't make up new ones.

%%%%%%%%%%%%%%%%%%%%%%%%%%%%%%%%%%%%%%%%%%%%%%%%%%

%%%%%%%%%%%%%%%%% BODY OF PAPER %%%%%%%%%%%%%%%%%%

\section{Introduction}\label{sec:intro}

The binary black holes (BBHs) detected by the Laser Interferometer
Gravitational-wave Observatory (LIGO) and the Virgo detectors
\citep{2020arXiv201014533T} are multiple times heavier than the black holes
(BHs) detected previously in X-ray binaries
\citep{mcclintock14,2016A&A...587A..61C}. Such a discrepancy has important
implications for the formation and evolution of BHs \citep[][and the references
therein]{2016ApJ...818L..22A,2020arXiv201014533T,2020ApJ...900L..13A}.  This
remarkable discovery also excited interests to find alternative
interpretations of the apparent high masses of the LIGO/Virgo BBHs. 

There are at least two astrophysical scenarios in which the mass of a
LIGO/Virgo BBH could be overestimated \citep[see][for a
review]{2020arXiv200907626C}.  Both scenarios are concerned with the effect of
``mass-redshift degenerate'', i.e., from gravitational waves (GWs) only a
redshifted mass $M(1+z)$ can be measured \citep{schutz86}, where $M$ is the
intrinsic mass of the source and $z$ is its redshift.  The first scenario is
related to gravitational lensing \citep{broadhurst18,smith18}.  In this case,
cosmological redshift causes the redshift of the mass, and lensing enhances the
amplitude of the detected GW, which in turn reduces the apparent (luminosity)
distance of the source. The large redshifted mass and the small apparent
distance conspire to result in an overestimation of the mass.  However, this
scenario has difficulty explaining all the heavy BBHs detected by LIGO/Virgo
mainly because the lensing rate is relatively low
\citep{hannuksela19,abbott19GWTC,2020arXiv201014533T}.  The second scenario is
concerned with the BBHs residing in the vicinity of supermassive black hole
(SMBH).  In this case, Doppler and gravitational redshifts predominate and
cause the observed high mass \citep{2019MNRAS.485L.141C}. To gain a large
redshift factor, the BBHs should form at a distance of a few Schwarzschild
radii ($r_s$) from the central SMBH.  One way of delivering BBHs to such a
small distance is tidal capture but it has been shown that the event rate is
also low \citep{addison15,2018CmPhy...1...53C}.

There is an alternative way of forming BBHs close to a SMBH.
It has been suggested that in active galactic nucleus (AGN), the accretion disc
surrounding the central SMBH may contain a large amount of compact objects (COs),
including stellar-mass BHs, neutron stars and white dwarfs.  On one hand, the
COs produced outside the accretion disc can be captured into it due to repeated
collisions with the disc \citep{1983ApJ...266..502N, 1991MNRAS.250..505S,
1993ApJ...409..592A,MacLeod20,melvyn20}.  On the other, massive stars can form in the outer,
gravitationally unstable part of the accretion disc and end up as COs
\citep{1980SvAL....6..357K, 1989ApJ...341..685S,
2004ApJ...608..108G,2007MNRAS.374..515L,wang10}.  Once in the accretion disc,
COs could form binaries efficiently by giving away excessive energy and angular
momentum to the surrounding gas \citep{2012MNRAS.425..460M,2019ApJ...878...85S,
2020ApJ...898...25T}.  Moreover, the interaction with the gas
\citep{2011ApJ...726...28B, 2017ApJ...835..165B, 2017MNRAS.464..946S} and the
other COs in the disc \citep{2018MNRAS.474.5672L,yang19}, or even the SMBH
\citep{antonini14}, also accelerates the coalescence of the binaries.  As a
result, the merger rate of the COs in AGN accretion discs could be significant.
For example, recent calculations suggest that the merger rate of the  BBHs in
AGN discs could be as high as $10-10^2 \, {\rm{Gpc}}^{-3} {\rm{yr}}^{-1}$
\citep{mckernan18,antoni19,2020ApJ...903..133S, grobner20,2020ApJ...898...25T}. 
Interestingly, it is reported that GW190521--the most massive LIGO BBH--has an
AGN counterpart \citep{2020PhRvL.124y1102G}.

If an AGN disc contains a large amount of BHs, it is natural to conjecture that
a fraction of them may migrate to a small distance of $\la10r_s$ from the
central SMBH due to hydrodynamical torque \citep[known as ``type I
migration'',][]{1979ApJ...233..857G, 1980ApJ...241..425G, 1991LPI....22.1463W,yunes11,kocsis13,barausse14}.
Consequently, BBHs may form there and  produce
seemingly overweight LIGO/Virgo BBHs, due to the large
Doppler and gravitational redshift \citep{2019MNRAS.485L.141C}.  However, an
earlier work shows that there are places outside a radius of  $10^2r_s$
where the positive and negative gas torques cancel out
\citep{2016ApJ...819L..17B}. As a result, the migration halts and BHs would be
trapped there. Although these ``migration traps'' are favorable places for BBH
formation \citep{2019ApJ...878...85S,yang19,2020ApJ...903..133S}, they are too
far from the central SMBH to induce any significant redshift effect.

Recently, \citet{2021arXiv210109146P} considered the effect of the thickness of
AGN accretion disc and found that the COs embedded in thick discs could
overcome the migration traps.  Their scenario is in close analogy with the
migration of planetesimals in protoplanetary discs. It is well known that 
protoplanetary discs are rotating slower than the Keplerian velocity because
they are partially supported by gas pressure \citep{1977MNRAS.180...57W}.
In particular, the deviation from Keplerian rotation increases
as the disc becomes thicker, because thicker discs have higher pressure.
Planetesimals, on the contrary, are
less affected by the gas pressure and are moving with Keplerian velocities. The
difference in velocity induces effectively a headwind on the planetesimals,
which drives the planetesimals to migrate inward \citep{1976PThPh..56.1756A,
1997Icar..126..261W, 1993prpl.conf.1031W, 2002ApJ...580..494Y}.  The headwind
mechanism also applies to AGN discs, and
according to \citet{2021arXiv210109146P} it is so efficient that the
BHs in the accretion disc always feel a negative torque so that they will quickly
migrate to
the innermost stable circular orbit (ISCO) of the central SMBH. 

The behaviour of COs close to the ISCO of AGN accretion disc has been studied
in the early works of
\citet{1993ApJ...411..610C,1996PhRvD..53.2901C,1994ApJ...436..249M}.
Interestingly, they revealed that in some cases the COs will, again, be
trapped.  It happens in a region close to the ISCO  where the gas rotates
faster than the Keplerian velocity. Such a super-Keplerian motion is the result
of a generic relativistic effect which causes the pressure gradient to point
inward to the SMBH \citep{abramowicz78,1988ApJ...332..646A,lancova19}. Since
now the COs rotate slower than the gas because their rotation velocity is
Keplerian, they will feel a tailwind and gain angular momentum. In the
original proposal of \citet{1993ApJ...411..610C}, the tailwind counteracts the
negative torque caused by the GW radiation of the CO-SMBH binary, and hence it
will halt the inward migration of the CO. We refer to this place as the ``last
migration trap'' because inside it there is no more place where COs can be
trapped.   

These previous works point to a picture that the COs in an AGN accretion disc
would be flushed, due to the headwind, to a place close to the ISCO and then be
trapped there, as a result of a flip of the direction of the wind. Because of
their closeness to the central SMBH, the GW sources forming in this last
migration trap will be affected by significant Doppler and gravitational
redshift.  As a first step towards understanding the merger rate of the COs
inside the last migration trap, in this paper we build an accretion disc model
which self-consistently generates the rotation velocity of the gas, especially
the transition from sub-Keplerian to super-Keplerian rotation. We analytically
study the migration of COs in such a disc and identify the parameter space
where the last migration trap exists. 

The paper is organized as follows. In Section~\ref{sec:method}, we describe our
disc model and lay out our scheme of calculating the drag force imposed on the
CO.  In Section~\ref{sec:result}, we derive the migration time-scale of a CO in
the aforementioned disc and look for the parameter space where the last
migration trap exists.  We discuss in Section~\ref{sec:discussion} the
limitation of our analytical method as well as the implications for future GW
observations.

\section{Models}\label{sec:method}

\subsection{Slim disc model}\label{subsec:model_slim}

The accretion disc of our interest should have a mild thickness, to induce a
non-Keplerian rotation. It should also have a relatively high density so as to
capture or form COs in it. Since we are interested in the region close to the
SMBH, we need to consider advective cooling because radiation becomes
insufficient to cool the disc.  These conditions are satisfied in the slim disc
model.  This model was constructed first assuming a pseudo-Newtonian potential
\citep{1988ApJ...332..646A} and later with general relativistic corrections
\citep{1996ApJ...471..762A, 1997ApJ...479..179A, 1998MNRAS.297..739B,
2009ApJS..183..171S, 2011arXiv1108.0396S}. Because of the relativistic
corrections, the transition of the inner disc from sub-Keplerian to
super-Keplerian rotation can be self-consistently modeled.  Since we need an
accurate disc profile extending to $r \lesssim 10 r_s$ to perform our
calculation, we adopt the full general relativistic disc model as is described
in Section~4 of \citet{2011arXiv1108.0396S}. 

The slim disc model adopted by us is governed by six equations. Four of them
are the same as the relativistic standard disc \citep{1973blho.conf..343N},
namely, the equations for mass conservation, angular momentum conservation,
vertical equilibrium and the equation of state. The fifth equation is for
energy conservation. Besides the normal terms of viscous heating and radiative
cooling, there is an extra term representing advection.  The last equation is
for momentum conservation in the radial direction.  It contains both the
pressure gradient and the radial velocity gradient. When they become large, the
actual angular velocity of the disc, $\Omega$, would deviate from the Keplerian
angular velocity, $\Omega_{\rm{k}}$. The departure from Keplerian motion gives rise
to the headwind and tailwind, which will act on the COs and affect their migration
in the accretion disc.

We simplify the disc model in two aspects so that we can perform the later
calculation relatively quickly. First, we assume that the central SMBH is not
spinning.  This assumption will increase the transition radius between
sub-Keplerian and super-Keplerian rotation, but the disc remains non-Keplerian
\citep{2009ApJS..183..171S,2011arXiv1108.0396S}.  Since the transition radius
determines the location of the last migration trap, our model would
overestimate the trapping radius for COs and subsequently lead to a
conservative estimation of the redshift effect.  The second simplification is
that we replace the vertical integration of gas density and pressure with a
simple multiplication of the height $H$ of the disc. Although this
simplification changes the corresponding terms by a constant coefficient, it
preserves the main feature of the slim disc, i.e., the disc remains thickness
and has a large pressure gradient.

There is no analytical solution to our slim disc model because there is an
eigenvalue, the specific angular momentum at the sonic point, which is not
known in {\it pri ori} and has to  be determined numerically
\citep{2011arXiv1108.0396S}.  In our work, we find this eigenvalue value by
integrating the disc out-side-in, starting from the outer boundary whose
physical conditions are set by the standard disc \citep{2008bhad.book.....K}.

Using this model, we can solve the disc structure for a SMBH in the mass range
of $M_{\rm{SMBH}} = 10^5- 10^9 \, M_{\odot}$ with an accretion rate of $
\dot{M} =  0.1 - 10 \, \dot{M}_{\rm{crit}}$, where $\dot{M}_{\rm{crit}}$ is the
critical accretion rate defined as $16 L_{\rm{Edd}} / c^2$
\citep{2011arXiv1108.0396S} with $L_{\rm{Edd}}$ being the Eddington luminosity.
We also try different values for the viscosity parameter $\alpha$, ranging from
$0.01$ to $0.1$. Given a set of parameters,  we solve the equations in the
radial range between $ r = \, 10^4 \, r_s$ and $3 \, r_s$. 

\begin{figure}
\centering
\includegraphics[width=0.5\textwidth]{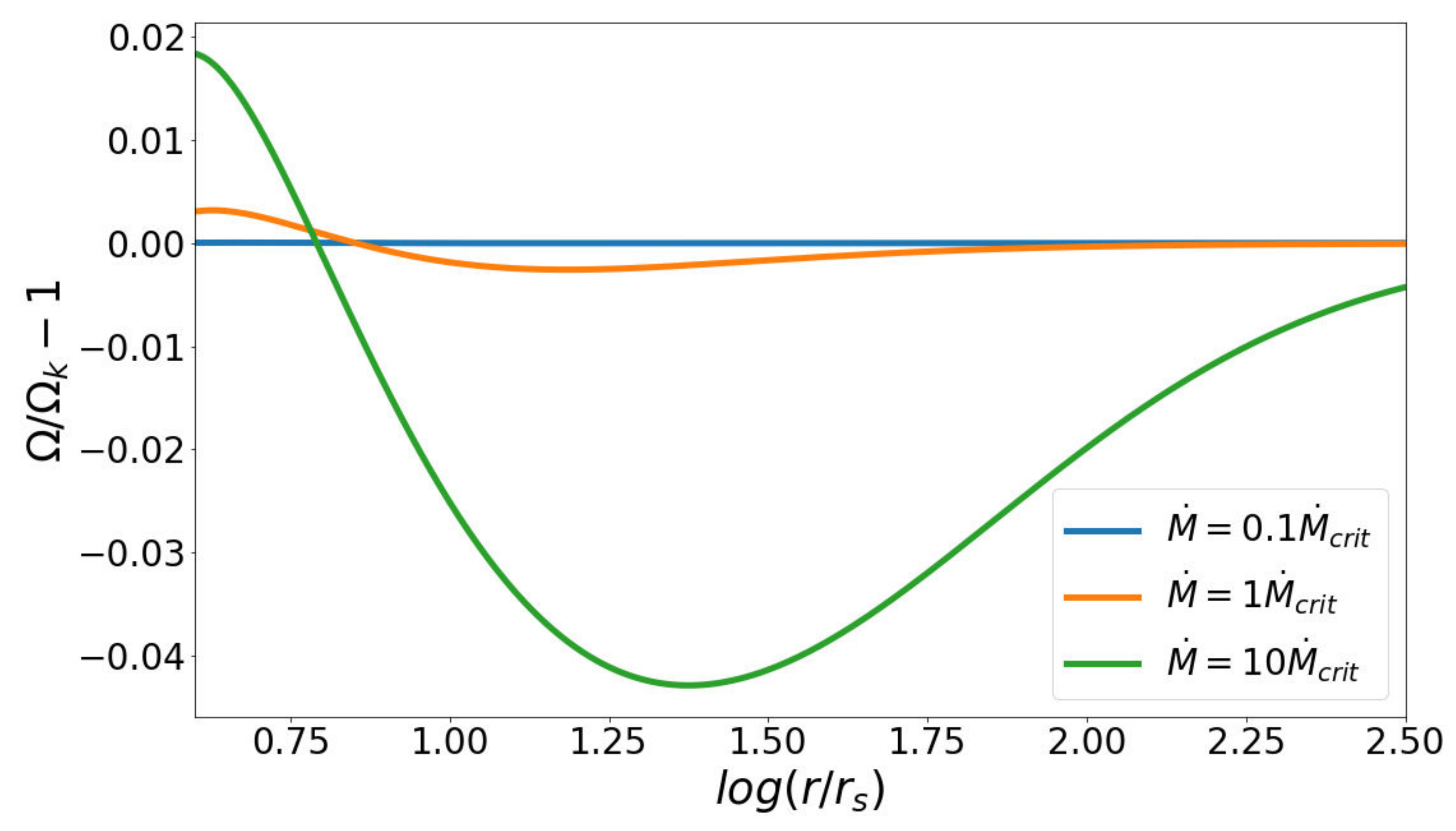}
\caption{The difference between the actual angular velocity of the gas $\Omega$
and the Keplerian angular velocity $\Omega_{\rm{k}}$ as a function of radius.
In this example, the mass of the SMBH is $10^8 M_{\odot}$ and the viscosity
parameter is $\alpha = 0.01$. The blue, orange and green lines correspond,
respective, to an accretion rate of $\dot{M} = $ 0.1, 1 and 10
$\dot{M}_{\rm{crit}}$. } \label{fig:omega_difference} 
\end{figure}

Figure~\ref{fig:omega_difference} shows the difference 
between the angular velocity derived from our slim disc model and the 
Keplerian angular velocity.  Here we fix the mass of the SMBH to
$M_{\rm{SMBH}} = 10^8 \, M_{\odot}$ as well as the viscous parameter to $\alpha
= 0.01$, and we vary the accretion rate between $\dot{M} =
0.1\dot{M}_{\rm{crit}}$ and $10\dot{M}_{\rm{crit}}$.  When $\dot{M} =  0.1 \,
\dot{M}_{\rm{crit}}$, the angular velocity is close to Keplerian throughout the
disc. With higher accretion rate, the deviation from the Keplerian velocity
becomes more prominent and could be several per cent at $r < 100 \, r_s$. For
all the three accretion rates, the discs are sub-Keplerian at $r \gtrsim 10 \,
r_s$ and super-Keplerian at $r \lesssim 10 \, r_s$.  Such a transition will
be important for our later study of the trapping radius of COs.  For other
values of $M_{\rm{SMBH}}$ and $\alpha$, the angular velocity profiles have a
similar behaviour and hence are not shown here.

\subsection{Drag force model}\label{subsec:model_force}

Unlike gas, COs around a SMBH are not significantly affected by the  radial
pressure gradient of the disc. Therefore, their rotation velocities remain
Keplerian.  Now that the COs are moving at a velocity different from the
velocity of the gas, they will feel a wind. In the sub-Keplerian region of the
accretion disc, such as the outer part of the disc, a CO will encounter a
headwind \citep{2021arXiv210109146P}, and in the super-Keplerian region, such
as the inner part of the disc, it will feel a tail wind
\citep{1993ApJ...411..610C}.  In either case, the wind would induce an
effective drag force on the CO \citep[see e.g.][for
discussion]{2021arXiv210109146P}. The difference is that the headwind drags the
CO backward and extracts angular momentum from the orbit, while the tailwind
pushes the CO forward and imparts angular momentum to the CO. Previous works
considered the headwind and tailwind separately
\citep{1993ApJ...411..610C,2021arXiv210109146P}, but here we will include both
effects in our model. The inclusion of both types of winds, as well as the
torque corresponding to GW radiation (see the next section), will allow us to
pin down the location where the net torque vanishes.  Such a location would be
our last migration trap.  

One essential ingredient that differentiates our model from the previous ones
is a mini accretion disc around the CO. Presence of the mini disc is a natural
consequence of the conservation of angular momentum, and earlier numerical
simulations suggest that its size could be as large as the Hills radius $R_{H} =
q^{1/3} r_0$  \citep{1976IAUS...73..237L, 1994ApJ...421..651A, 2011ApJ...726...28B}, where $q =
M_{\rm{CO}}/M_{\rm{SMBH}}$ is the mass ratio between the CO and the SMBH, and
$r_0$ denotes the orbital radius of the CO around the SMBH.  The
mini disc significantly enhances the interaction cross section of the CO with
the gas in the AGN disc, because any fluid parcel crossing the mini disc would
exchange momentum with the CO. 
For example, in the conventional scenario of Bondi-Hoyle-Lyttleton accretion
the cross section of exchanging momentum is determined by 
$\Sigma_{\rm BHL}\sim G^2M^2_{\rm CO}/ (V^2_{\rm rel}+c_s^2)^2$, where 
$V_{\rm{rel}}$ is the relative velocity between the CO and the gas
in the surrounding AGN disc, and $c_s$ is the sound speed of the gas.
Comparing it to the physical cross section of the Hills sphere, 
$\Sigma_H\sim R^2_{H}$, we find that
\begin{equation} \label{eq: cross section ratio}
	\frac{\Sigma_H}{\Sigma_{\rm BHL}}\ga
	\left(\frac{M_{\rm SMBH }}{M_{\rm CO}}\right)^{4/3}
	\left(\frac{H}{r_0}\right)^{4},
\end{equation}
where we have used the condition of hydrostatic equilibrium, $c_s\simeq \Omega_{\rm{k}} (H/r_0) $.
In this work, typically, $(H/r_0) \sim 0.1 $ for $\dot{M} =  1 \dot{M}_{\rm{crit}}$.
So the above ratio is significantly greater than unity in the slim disc region where
$H$ is not much smaller than $r_0$. This result indicates that we have to consider the
interaction between the mini disc and the fluid in the AGN accretion disc so as to
properly account for the drag force imposed on the CO.  

Following the above scenario, we now calculate the drag force exerted on the mini disc.
The CO feels the same force because of a strong gravitational coupling between the
mini disc and the CO \citep[e.g. see][for a discussion]{chen20PR}.
Without loss of generality, we write the momentum flux imparted onto the
mini disc as 
\begin{equation}\label{eq:force_1}
    F_{\rm{gas}}  = k\rho(r_0) R^2_H V_{\rm{rel}}^2,
\end{equation}
where $\rho(r_0)$ is the unperturbed gas density in the AGN accretion disc,
evaluated at the orbital radius $r_0$ of the CO. The geometrical coefficient
$k$ depends on the actual velocity field of the gas around the CO and will be
determined later. In principle it is of order unity because the size of the
Hills sphere sets a characteristic cross section for gas-disc
interaction.  The direction of the drag force is the same as the velocity of
the gas relative to the CO, because it is the direction where
momentum is added to the mini disc.  Note that using our notation, the relative
velocity is $V_{\rm{rel}} = \left[\Omega(r_0)-\Omega_{\rm{k}}(r_0)\right]r_0$.

\begin{figure*}
\begin{minipage}[t]{0.45\linewidth}
\centering
\includegraphics[width=2.9in]{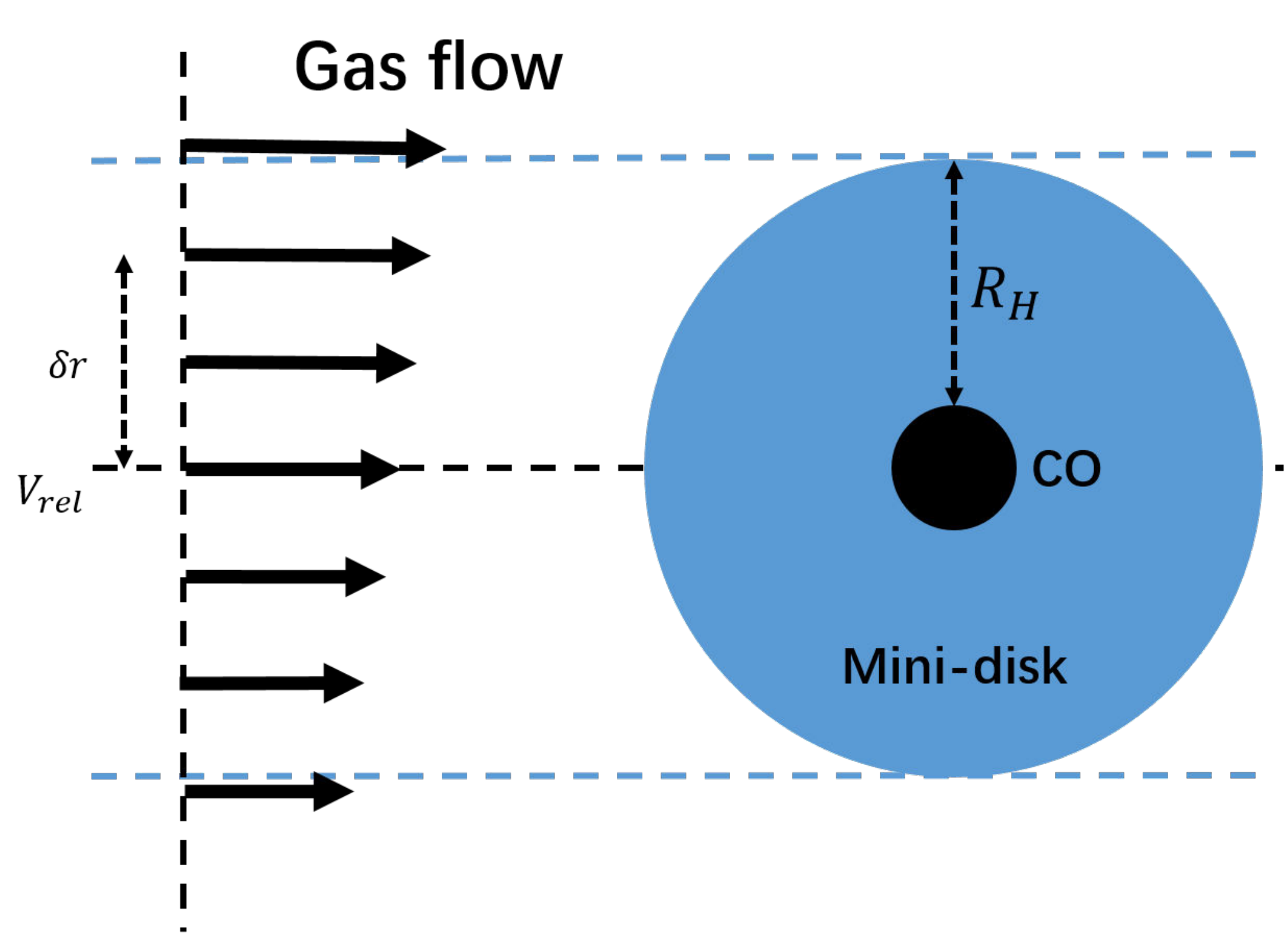}
\end{minipage}
\begin{minipage}[t]{0.45\linewidth}
\centering
\includegraphics[width=3in]{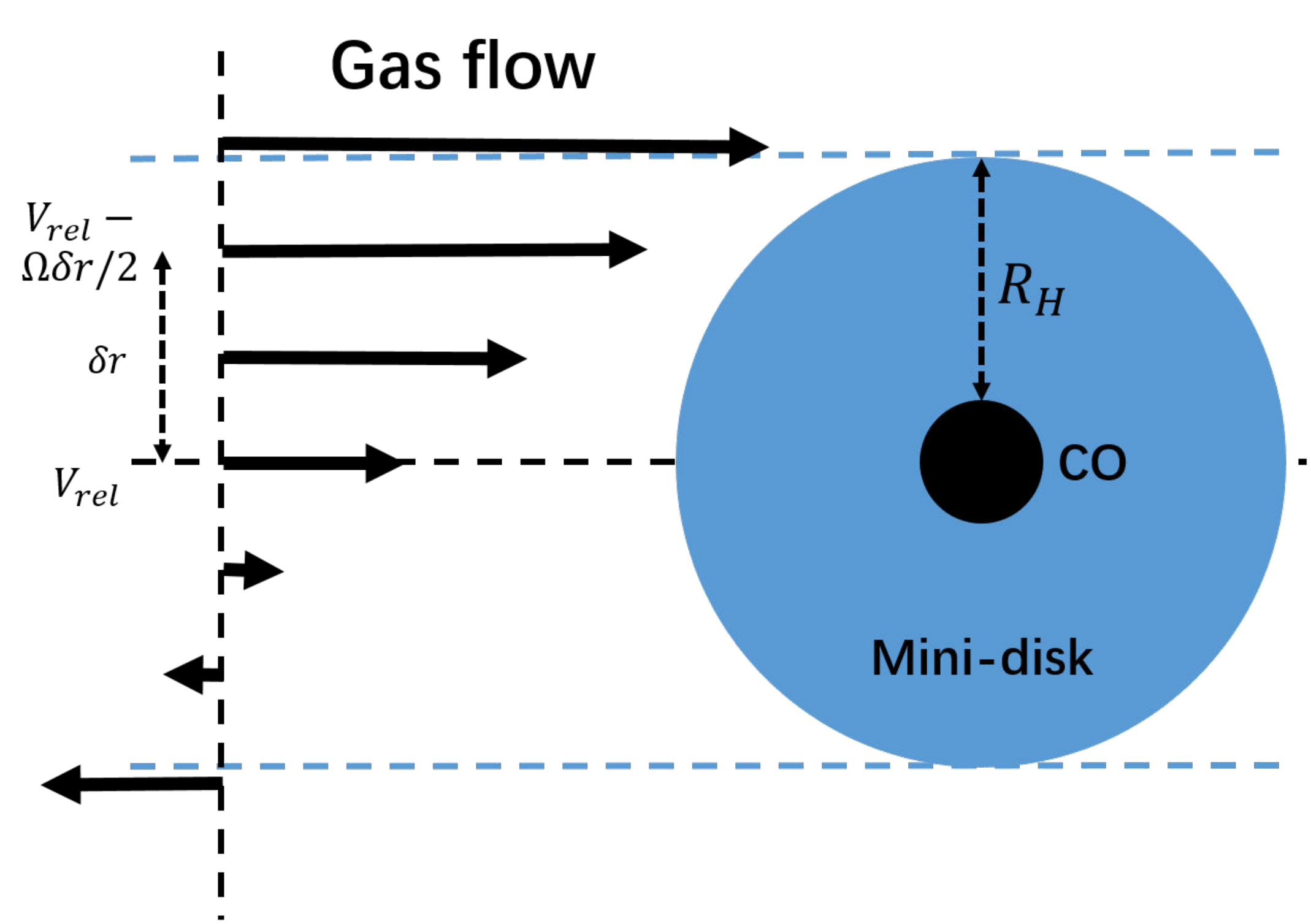}
\end{minipage}
\caption{Sketch of the mini-disc of a CO and the velocity field of 
the incoming gas flow. The vertical direction coincides with the radial
direction of the AGN disc. The horizontal arrows indicate the direction and magnitude
of the relative velocity between the gas flow and the CO.
The left panel shows the case in which the differential
rotation of the AGN accretion disc is small, so that the velocity of the gas
relative to the CO is more or less constant across the surface of the mini
disc. This case corresponds to a large distance between the CO and the SMBH.
The right panel shows the case where the differential
rotation of the AGN accretion disc is significant. The CO in this case is close
to the SMBH.
}
\label{fig:differential rotation}
\end{figure*}

To determine the geometrical factor $k$, we have to first understand the
velocity field of the gas around the CO. We notice that previous works only
considered a single value for the relative velocity between the CO and the gas
\citep{1993ApJ...411..610C,2021arXiv210109146P}, which is the aforementioned
$V_{\rm rel}$.  This approximation is valid in the region where differential
rotation of the AGN accretion disc is negligible (see the left panel of
Figure~\ref{fig:differential rotation}).  In our case, since the CO is
relatively close to the SMBH ($\la10^2r_s$) and the differential rotation
velocity increases with decreasing $r_0$, we have to consider a variation of
$V_{\rm rel}$ across the characteristic scale of the mini disc.  Now consider a
fluid element in the AGN accretion disc which has an initial orbital radius
$\delta r + r_0$, where $|\delta r|\ll r_0$ for our problem. Its orbital
velocity is $(\delta r + r_0) \Omega(\delta r + r_0) \approx r_0 \Omega(r_0) +
d [ r\Omega(r)]/dr \, \delta r \approx \Omega(r_0) r_0 - \Omega(r_0) \delta r /
2 $, where we have assumed that $\Omega(r)$ does not significantly deviate from
$\Omega_{\rm{k}}(r)$, which is valid according to
Figure~\ref{fig:omega_difference}.  Therefore, the velocity relative to the CO
is $ V_{\rm rel}(r_0) - \Omega(r_0) \delta r /2$, where the first term comes
from the non-Keplerian motion of the fluid and the second term from the
differential rotation.  As is illustrated in the right panel
of Figure~\ref{fig:differential rotation}, the differential rotation could even
flip the direction of $V_{\rm rel}$ across a radial scale of $R_H$ if $r_0$ is
small. Consequently, different parts of the mini disc feels different types of
wind. The headwind and tailwind, combined, give rise to a net force acting on
the CO.

Knowing the velocity field of the gas, we now proceed
to determine which part could enter the Hills radius and fuel the mini disc.
Here we cannot use the standard formula for gravitational focusing because
the background is not flat--it is determined by the tidal field of the SMBH.
Instead, we use the ``impulse model'' \citep{2010apf..book.....A}.
According to this model, when a fluid element
passes by the CO, the gravitational force of the CO on the fluid element is the strongest
within the time interval $\delta t = 2 |\delta r| / | V_{\rm rel}(r_0) -
\Omega(r_0) \delta r /2 | $. During this time interval, the velocity of the
fluid element changes by $ \delta v\sim \delta t G M_{\rm{CO}}/\delta r^2$. Such a
velocity change will excite the orbital eccentricity of the fluid element 
to a value of $\delta v/(\Omega r)$. If this orbital
eccentricity falls in the range of $ | (\delta r \mp R_{\rm{H}} ) / r_0 | $,
the fluid element could reach the mini disc and is in the so-called 
``feeding zone''. Following this
argument, we find that the boundaries of the feeding zone satisfy the equation 
\begin{equation}
     4 G M_{\rm{CO}}  = \pm \Omega^2  \delta r^3  \mp 2 V_{\rm{rel}} \Omega \delta r^2 - R_{\rm{H}} \Omega^2 \delta r^2 + 2 R_{\rm{H}} V_{\rm{rel}} \Omega \delta r,\label{eq:boundary}
\end{equation}
where plus and minus signs correspond to gas moving in opposite directions
relative to the CO. One can verify the above equation by considering a
Keplerian disc, i.e., in the case where the second and fourth terms on the
right hand side vanish.  One will recover the result that the boundaries of the
feeding zone are at $\delta r_{\pm} = \pm 2 R_{\rm{H}}$
\citep{2010apf..book.....A}.  For our non-Keplerian disc, the second and fourth
terms on the right hand side of Equation~(\ref{eq:boundary})  do not vanish.
However, they are small relative to the first and third terms so that  we can
expand $\delta r $ to the order of $V_{\rm{rel}} / \Omega(r_0)$ and derive
$\delta r_{\pm} = \pm 2 R_{\rm{H}} +  V_{\rm{rel}} /2 \Omega(r_0)$.  Not all
the gas within the above boundary can interact with the mini disc. Some fluid
elements are on the so-called  ``horseshoe orbit'' which does not enter the
Roche radius of the CO \citep{1991LPI....22.1463W}. Here we follow the
treatment used in \citet{2010apf..book.....A} and assume that the width of the
feeding zone is half of the width of the boundaries. 

Now we know the width of the feeding zone ($|\delta r_{\pm}/2|$)
and the velocity field in it ($V_{\rm rel}(r_0) - \Omega(r_0) \delta r /2$),
we can calculate the momentum flux deposited onto the mini disc and derive the drag force
exerted on the CO.
Assuming that the height of the feeding zone is $|\delta r_{\pm}/2|$ \citep{2010apf..book.....A}, 
we find that
the drag force (to the lowest order of
$V_{\rm{rel}} / \Omega R_{\rm{H}}$) is
\begin{eqnarray}
    F_{\rm{gas}} &=& - \rho(r_0) ( V_{\rm{rel}} - \Omega(r_0) \delta r_{+} /2 )^2 {\delta r_{+}}^2 /4  \nonumber \\
   && + \rho(r_0) ( V_{\rm{rel}} - \Omega(r_0) \delta r_{-} /2 )^2 {\delta r_{-}}^2 /4  \\
     &=&  - \rho \Omega^2(r_0) [ R_{\rm{H}} - 3 V_{\rm{rel}}/4\Omega(r_0) ] ^2 [ R_{\rm{H}} + V_{\rm{rel}}/4\Omega(r_0) ] ^2  \nonumber \\
     &+& \rho \Omega^2(r_0) [ R_{\rm{H}} + 3 V_{\rm{rel}}/4\Omega(r_0) ] ^2 [ R_{\rm{H}} - V_{\rm{rel}}/4\Omega(r_0) ] ^2   \\
     &\sim& \rho R_{\rm{H}}^3 \Omega(r_0) [\Omega(r_0) -\Omega_{\rm{k}}(r_0)] r_0.\label{eq:force_2}
\end{eqnarray}
The direction of the force is also the same as $V_{\rm rel}(r_0)$. Therefore, the
CO feels a headwind (tailwind) if it resides in the  
sub-Keplerian (super-Keplerian) region of the disc. 

We can unify the Equations~(\ref{eq:force_1}) and (\ref{eq:force_2}) by
writing the drag force as
\begin{equation}
F_{\rm{gas}}= k \rho(r_0) R_{\rm{H}}^2 [\Omega(r_0)-\Omega_{\rm{k}}(r_0)]^2 r^2_0 \, , \label{eq:force_3}
\end{equation}
where $k=\max\left[ 1 , \, \Omega(r_0) R_{\rm{H}} /
(|\Omega(r_0)-\Omega_{\rm{k}}(r_0)| r_0)  \right]$.  We can see that when the
differential rotation is unimportant, i.e., $\Omega(r_0) R_{\rm{H}} <
|\Omega(r_0)-\Omega_{\rm{k}}(r_0)| r_0 $, we have $k = 1$ and we recover
Equation~(\ref{eq:force_1}).  On the contrary, if the differential rotation is
not negligible, i.e.,  $\Omega(r_0) R_{\rm{H}} >
|\Omega(r_0)-\Omega_{\rm{k}}(r_0)| r_0 $, we have $k = \Omega(r_0) R_{\rm{H}} /
|\Omega(r_0)-\Omega_{\rm{k}}(r_0)| r_0 $ and Equation~(\ref{eq:force_2}) is
recovered.

We note that the derivation of the drag force assumes that $R_H<H$.  This
assumption is valid because in our model $R_H/H=(r_0/H)(M_{\rm CO}/M_{\rm
SMBH})^{1/3}$.  Note that the thickness of the disc is related to the
non-Keplerian motion of the disc (see Figure~\ref{fig:omega_difference}) as
\begin{equation}
	\left(\frac{H}{r_0}\right)^2 \simeq 2 \left| \frac{\Omega}{\Omega_{k}}-1 \right|
	\label{eq:height}
\end{equation} 
according to Equations (4.7), (4.10) and (4.12) of \citet{2011arXiv1108.0396S}
in the non-relativistic limit, which, in turn, is related to the accretion
rate.  For example, we find that $H/r \sim 0.01, \, 0.1, \, 0.3$ at $r < 100 \,
r_s$, when the accretion rate is $\dot{M} =  0.1, \, 1, \, 10 \,
\dot{M}_{\rm{crit}}$. As a result, the value of $R_H/H$ is smaller than $1$ in
the region of $r_0\la10^2r_s$ around the SMBHs of our interest ($M_{\rm
SMBH}\ga10^7M_\odot $).

We also note that we can neglect the drag force induced by the 
radial advection of the slim disc because it is much weaker than 
the centrifugal force. The force exerted by the radial advection
on the mini disc can be estimated with $\rho R_H^2
V_r^2$, where $V_r$ is the radial velocity of the gas. The centrifugal force is
$M_{\rm{CO}} \Omega_{\rm{k}}^2 r_0$. The ratio of the two forces is:
\begin{equation}
    \frac{\rho R_H^2 V_r^2}{M_{\rm{CO}} \Omega_{\rm{k}}^2 r_0} = q^{-1/3} \frac{V_r^2}{\Omega_{\rm{k}}^2 r_0^2} \frac{\rho r_0^3}{M_{\rm{SMBH}}} \sim q^{-1/3} \alpha^2 (\frac{H}{r_0})^3 \frac{\Sigma r_0^2}{M_{\rm{SMBH}}} ,
\end{equation}
where $\Sigma \sim H \rho$ is the surface density of the disk and we have
used $V_r=\alpha H^2 \Omega_{\rm{k}}^2 $ based on the equations of mass conservation and
angular momentum conservation. With typical parameters in this work
($q \geq 10^{-8}$, $\alpha \leq 0.1$, $H/r_0 \lesssim 0.3$, $\Sigma
r_0^2/M_{\rm{SMBH}} \lesssim 10^{-4}$), we find that the drag forced induced by the
radial advection is negligible.

\section{Results}\label{sec:result}

Given the disc and drag force models, we can study the migration of COs in our
non-Keplerian disc.  We first compare four characteristic time-scales which
outline the behavior of a CO in AGN disc.  These four time-scales are (i) the
migration time-scale $T_{\rm{gas}}$ due to the interaction with the
headwind/tailwind, (ii) type I migration time-scale $T_{I}$ as has been studied
extensively in the previous works on COs in AGN discs
\citep[e.g.][]{2016ApJ...819L..17B}, (iii) orbital decay time-scale of the CO
due to GW, $T_{\rm GW}$, which becomes particularly important at small radius
$r_0$ and (iv) the lifetime of the AGN $T_{\rm AGN}$.  In general,
the shortest time-scale of $T_{\rm{gas}}$, $T_{I}$ and $T_{\rm GW}$ determines
which mechanism drives the migration of COs.  However, if $T_{\rm AGN}$ becomes
the shortest time-scale, the accretion disc would disappear and the migration
of CO would stall. In this case a CO cannot reach a close distance to the central
SMBH.

To calculate $T_{\rm gas}$, we use Equation~(\ref{eq:force_3}) and write
\begin{equation}
    T_{\rm{gas}} = \frac{M_{\rm{CO}} \Omega_{\rm{k}} r}{F_{\rm{gas}}} = k^{-1} q^{1/3} \frac{ \Omega_{\rm{k}}^2}{(\Omega-\Omega_{\rm{k}})^2}  \frac{ M_{\rm{SMBH}} }{\rho r^3} \frac{1}{\Omega_{\rm{k}}} \, .
\end{equation}
Using the relationship derived in Equation~(\ref{eq:height}) and the approximation that
$\rho\sim\Sigma/H$, we can derive a more quantitative equation for $T_{\rm gas}$ as  
\begin{eqnarray}
	\nonumber T_{\rm{gas}}  &=& 4 k^{-1} q^{\frac{1}{3}} \left(\frac{H}{r}\right)^{-3} \frac{ M_{\rm{SMBH}} }{\Sigma r^2} \frac{1}{\Omega_{\rm{k}}} = 0.18 {\rm{Myr}} \,  k^{-1} \left(\frac{M_{\rm{CO}}}{ 10 M_{\odot}}\right)^{\frac{1}{3}}  \\
	&\times& \left(\frac{M_{\rm{SMBH}}}{ 10^8 M_{\odot}}\right)^{-\frac{4}{3}} 
	\left(\frac{r}{10 r_s}\right)^{-\frac{3}{2}}  
	\left(\frac{H/r}{0.1}\right)^{-3} 
	\left(\frac{\Sigma}{10^{5} \, {\rm{g}} \, {\rm{cm}}^{-2}}\right)^{-1}. 
     \label{eq:T_gas}
\end{eqnarray}
In the last equation we have chosen the typical values of $M_{\rm{CO}}$,
$M_{\rm{SMBH}}$ and $\Sigma$ for scaling. We note that depending on the direction of the wind, the CO
would migrate either inward or outward on the time-scale of $T_{\rm gas }$

Type I migration is caused by the interaction between the CO and the density waves in the disc
\citep{1980ApJ...241..425G}.
The direction of the migration, as is noted in the earlier work of migration trap \citep{2016ApJ...819L..17B},  could be either inward or outward depending on 
the exact temperature and density profiles of the accretion disc. We follow
\citet{1997Icar..126..261W} and calculate the corresponding time-scale as
\begin{equation}
    T_{I}  = q^{-1}  \frac{M_{\rm{SMBH}}}{\Sigma r^2}  \frac{1}{\Omega_{\rm{k}}}.
\end{equation}
Compare it with the time-scale due to gas drag, we find that
\begin{eqnarray}\label{eq:time-scale_comparison_1}
     T_{\rm{gas}}/T_{I} = 4 k^{-1} q^{1/3} (\frac{H}{r})^{-3} \frac{ M_{\rm{SMBH}} }{\Sigma r^2} \frac{1}{\Omega_{\rm{k}}} \bigg/  q^{-1}  \frac{M_{\rm{SMBH}}}{\Sigma r^2}  \frac{1}{\Omega_{\rm{k}}} \\
    = 4 q^{4/3} {(H/r)}^{-5} = 0.6 (\frac{M_{\rm{CO}}}{ 10 M_{\odot}})^{4/3} (\frac{M_{\rm{SMBH}}}{ 10^8 M_{\odot}})^{-4/3} (\frac{H/r}{0.02})^{-5}.
\end{eqnarray}
Apparently, type I migration becomes less efficient than the gas drag as the
mass of the CO decreases. Besides, a heavier SMBH or a thicker accretion
disc could also suppress the relative importance of type I migration. For these
reasons, in the following analysis we focus on the cases in which
$M_{\rm{SMBH}} \gtrsim 10^7 M_\odot$, $M_{\rm{CO}} \lesssim 100 M_\odot$ and
$H/r \gtrsim 0.01$. The last condition about the disc scale height corresponds
to an accretion rate of $\dot{M} \gtrsim 0.1 \dot{M}_{\rm{crit}}$. 

As the CO is driven closer to the SMBH, the GW radiation of the SMBH-CO system becomes 
more powerful and needs to be accounted for in the calculation of the migration time-scale.   
The orbital decay time-scale due to GW radiation can be calculated with
\begin{equation}
    T_{\rm{GW}} = \frac{5c^5 r^4}{256G^3 (M_{\rm{SMBH}} + M_{\rm{CO}}) M_{\rm{SMBH}} M_{\rm{CO}} }
\end{equation}
\citep{1964PhRv..136.1224P}. It follows that
\begin{eqnarray}
 \frac{T_{\rm{gas}}}{T_{\rm{GW}}} &=& \frac{128 \sqrt{2}}{5} q^{4/3} (1+q) 
\frac{ M_{\rm{SMBH}} }{\Sigma r^2} 
\left(\frac{H}{r}\right)^{-3} 
\left(\frac{r}{r_s}\right)^{-5/2}\\   
&=& 1.2 \left(\frac{M_{\rm{CO}}}{ 10\, M_{\odot}}\right)^{\frac{4}{3}} 
\left(\frac{M_{\rm{SMBH}}}{ 10^8 \, M_{\odot}}\right)^{-\frac{10}{3}} 
\left (\frac{r}{10 \, r_s}\right)^{-\frac{11}{2}} \nonumber\\ 
&\times&\left(\frac{H/r}{0.1}\right)^{-3} 
\left(\frac{\Sigma}{10^5 \, {\rm{g}} \, {\rm{cm}}^{-2}}\right)^{-1} .
\label{eq:time-scale_comparison_2}
\end{eqnarray}
Equation~(\ref{eq:time-scale_comparison_2}) shows that at a close distance
$r\sim 10r_s$, where the AGN accretion disc is thick and super-Keplerian, the
migration time-scale due to gas drag $T_{\rm gas}$ could be comparable to the GW
radiation time-scale $T_{\rm GW}$, provided that the CO is small and the SMBH is
massive.  In this case, the gas gives angular momentum to the CO at a rate
comparable to the rate the GW radiation extracts angular momentum from the CO.
As a result, the CO stalls at such a small radius. 

Such a stalling radius is absent in the work of 
\citet{2021arXiv210109146P}, because they did not use the slim disc model
so that their disc lacks a super-Keplerian inner region.
Our stalling radius is also different from what have been found by
\citet{1993ApJ...411..610C,1994ApJ...436..249M} because
we consider a mini disc surrounding the CO. The mini disc taps more angular momentum
from the tailwind, and hence
results in a smaller
stalling radius than what \citet{1993ApJ...411..610C,1994ApJ...436..249M} have predicted.

The lifetime of AGN determines how long an accretion disc is present around the
SMBH. We are interested in a particular phase of AGN during which the accretion
rate is higher than $\dot{M} \gtrsim 0.1 \dot{M}_{\rm{crit}} $. We denote the
duration of such a phase as $T_{\rm AGN}$.  After this phase, the disc is
geometrically thin and the super-/sub-Keplerian motion of the gas becomes less
prominent.  However, it is challenging to predict the value of $T_{\rm AGN}$
\citep{2009ApJ...690...20S}. This is so because observations show that AGNs
vary significantly on different time-scales, from several days
\citep{2017ApJ...834..111C} to less than 1 Myr \citep{2015MNRAS.451.2517S,
2018MNRAS.474.2444S}, and even as long as several hundred Myrs
\citep{2008ApJ...676..816G}. In this work, for simplicity we assume that
$T_{\rm AGN}= 10^7$ yrs. It is a typical value for high luminosity AGNs
\citep{2008ApJ...676..816G,2009ApJ...698.1550H,2013MNRAS.434..606G}. Since it
is shorter than the Salpeter time-scale $   T_{\rm{Sal}} = 3 \times 10^7
(\dot{M}_{\rm crit}/\dot{M}) \rm{yrs}$ \citep{1964ApJ...140..796S} unless
$\dot{M}$ is significantly greater than $3\dot{M}_{\rm crit}$, we will neglect
the variation of the mass of the SMBH in our model.

For massive COs, 
type II migration also becomes important \citep{1986ApJ...309..846L}. This type of
migration requires that the CO opens a gap in the AGN accretion disc due to
tidal interaction. After the gap forms, the migration of the CO is driven
mainly by the torque exerted by the density waves at the edges of the gap.  For
a CO in a Keplerian thin disc, it can be shown that the type II migration
time-scale is short, of the order of $T_{\rm Sal}(M_{\rm CO}/M_{\rm SMBH})$
\citep[e.g.][]{chen20PR}.  Therefore, the CO could migrate inward efficiently.
For slims discs, however, radial advection becomes important in the transportation
angular momentum, but there is no analytical or numerical calculation of the gap
formation and migration time-scales. For this reason, in this work we do not
consider the COs undergoing type II migration. In practice, we check the criteria 
for gap opening
\begin{equation} \label{eq: crit2}
    \frac{R_{\rm{H}}}{H} (\frac{8 H/r}{27 \pi \alpha})^{1/6} > 1 \quad {\rm{or}} \quad \frac{R_{\rm{H}}}{H} (\frac{H/r}{1600 \alpha})^{1/3} > 1.
\end{equation}
\citep{1986ApJ...309..846L, 1999ApJ...514..344B}, and stop the calculation when
these criteria are met. 

Although the time-scales $T_{\rm gas}$, $T_I$ and $T_{\rm GW}$ derived above
are informative about the efficiencies of different migration types, they do
not show the directions of the migration.  Therefore, we further derive the
torque exerted on the CO, to show both the magnitude and the direction of the
corresponding effect. In the following, a positive torque drives the CO outward
and a negative one drives it inward.  The torque due to the headwind/tailwind
can be written as $\Gamma_{\rm{gas}} = F_{\rm{gas}} r$, where the direction is
determined by the sign of the force $F_{\rm gas}$.  For Type I migration, the
torque $\Gamma_{I}$ is calculated as in \citet{2010MNRAS.401.1950P}.  For GW
radiation, the torque $\Gamma_{\rm{GW}}$ is always negative and has been derived in
\citet{1964PhRv..136.1224P}. The sum of the torques, $\Gamma_{\rm{sum}} =
\Gamma_{\rm{gas}} + \Gamma_{I} + \Gamma_{\rm{GW}}$, determines the final
direction of the migration.

\begin{figure}
\centering
\includegraphics[width=0.5\textwidth]{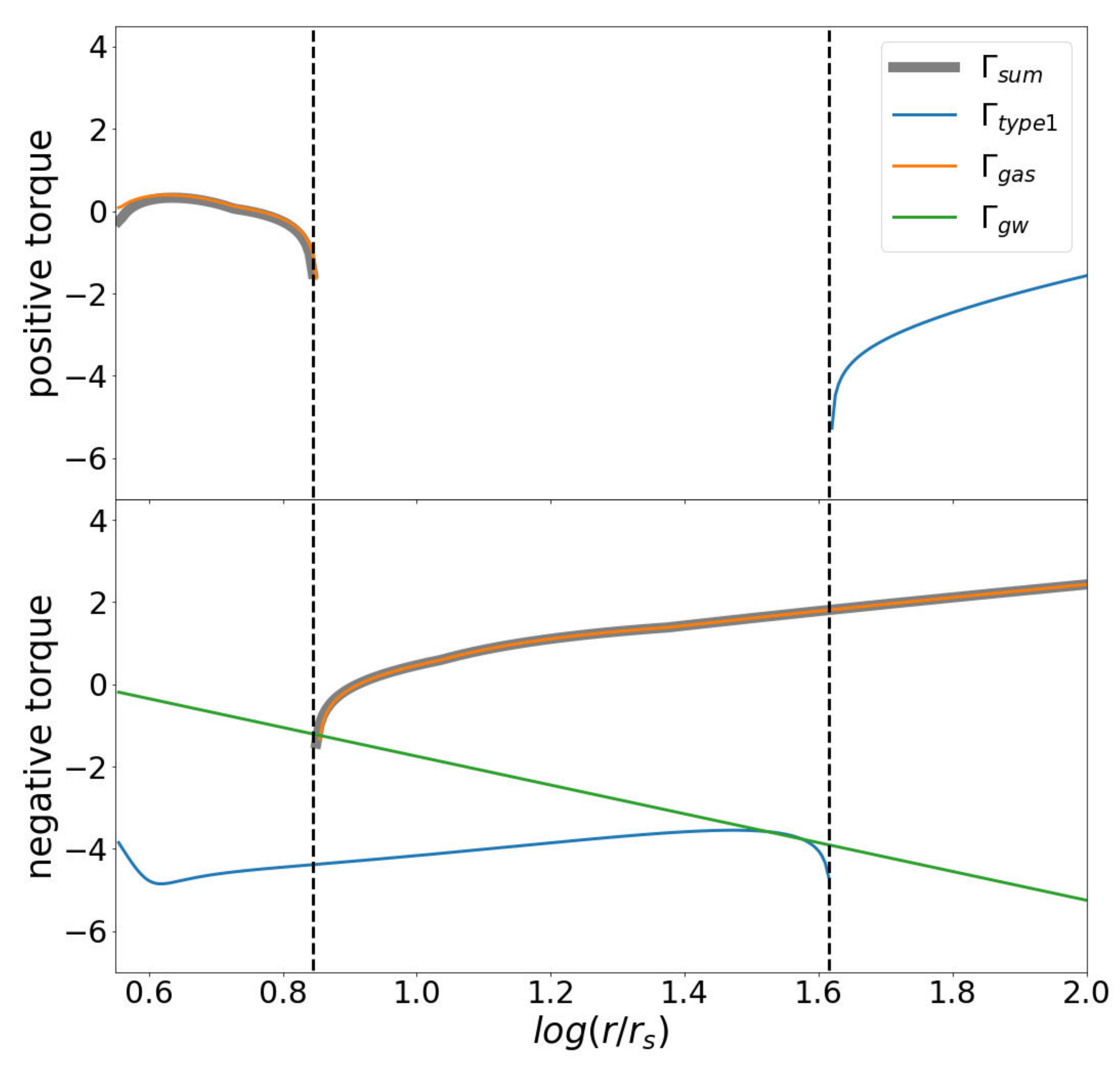}
\caption{Different types of torques as a function of the distance from the
central SMBH.  The blue, orange, green and grey curves represent the torques
due to Type I migration, headwind/tailwind, GW radiation and the sum of the
previous three. The unit of the torque is $10^{-10} M_{\rm{CO}} c^2$ and the radius
is normalized by the Schwarzschild radius of the SMBH $r_s$. The upper panel shows the positive torques, which
drive the COs outward, while the lower panel shows the negative ones, which
drive the COs closer to the SMBH.  The dashed vertical lines mark the locations
where a torque changes sign.  In this example the parameters are set to
$M_{\rm{CO}} = 10 M_{\odot}$, $M_{\rm{SMBH}} = 10^{9} M_{\odot}$,
$\dot{M}=1\dot{M}_{\rm{crit}}$ and $\alpha = 0.01$.  } \label{fig:result1}
\end{figure}

Figure~\ref{fig:result1} shows the values of $ \Gamma_{\rm{gas}} , \,
\Gamma_{I} , \, \Gamma_{\rm{GW}}$ and $\Gamma_{\rm{sum}}$ at different radius
of the AGN accretion disc. In the calculation we have assumed that $M_{\rm{CO}} = 10
M_{\odot}$, $M_{\rm SMBH}=10^{9} M_{\odot}$ and $\dot{M}=1\dot{M}_{\rm{crit}}$.
These torques are normalized by $10^{-10} M_{\rm{CO}} c^2$. The upper panel
shows the torques which drive the COs outward and the lower one shows the
torques which drive the COs closer to the SMBH. In both panels, we find that
the torque $\Gamma_{\rm{gas}}$ due to headwind/tailwind predominates.  In
particular, at $ r \gtrsim 30 \, r_s$, the negative gas drag torque (orange
curve) overcomes the positive, type I migration torque (blue curve).  This
result indicates that the headwind
could push the COs out of the migration traps and force them to migrate closer
to the SMBH.  The same result was derived recently by
\citet{2021arXiv210109146P}, who used a different accretion disc model and did
not consider the effects of the mini disc. Our result complements theirs and
supports their conclusion that the type I migration traps as is envisioned in
\citet{2016ApJ...819L..17B} may not exist in thick discs. 

However, Figure~\ref{fig:result1} reveals a new kind of migration trap at a
small radius of $r \sim 7 r_s$ (see the dashed vertical line on the left). Here
the gas drag (orange curve) still predominates. Nevertheless, the total torque
(orange curve) changes sign because the accretion flow transforms from
sub-Keplerian to super-Keplerian, i.e., the wind direction changes. COs could
be trapped here because at $r \ga 7 r_s$ the total torque is negative, forcing the
COs to migrate inward,  and at $r \la7 r_s$ the total torque becomes positive,
forcing the COs to migrate outward. This trap is characteristically different
from the type I migration trap because the driving force is the
headwind/tailwind from the non-Kepler disc. It is also different from the
stalling mechanism  found by \citet{1993ApJ...411..610C,1994ApJ...436..249M},
because they considered a balance between the GW torque and the torque due to
tailwind, but here we find that the GW torque is much weaker than those due to
headwind/tailwind. Most importantly, inside the radius of $r \simeq 7 r_s$ there is no more
place where the total torque changes sign. So the location
$r \simeq 7 r_s$ is indeed ``the last migration trap''.

\begin{figure}
\centering
\includegraphics[width=0.5\textwidth]{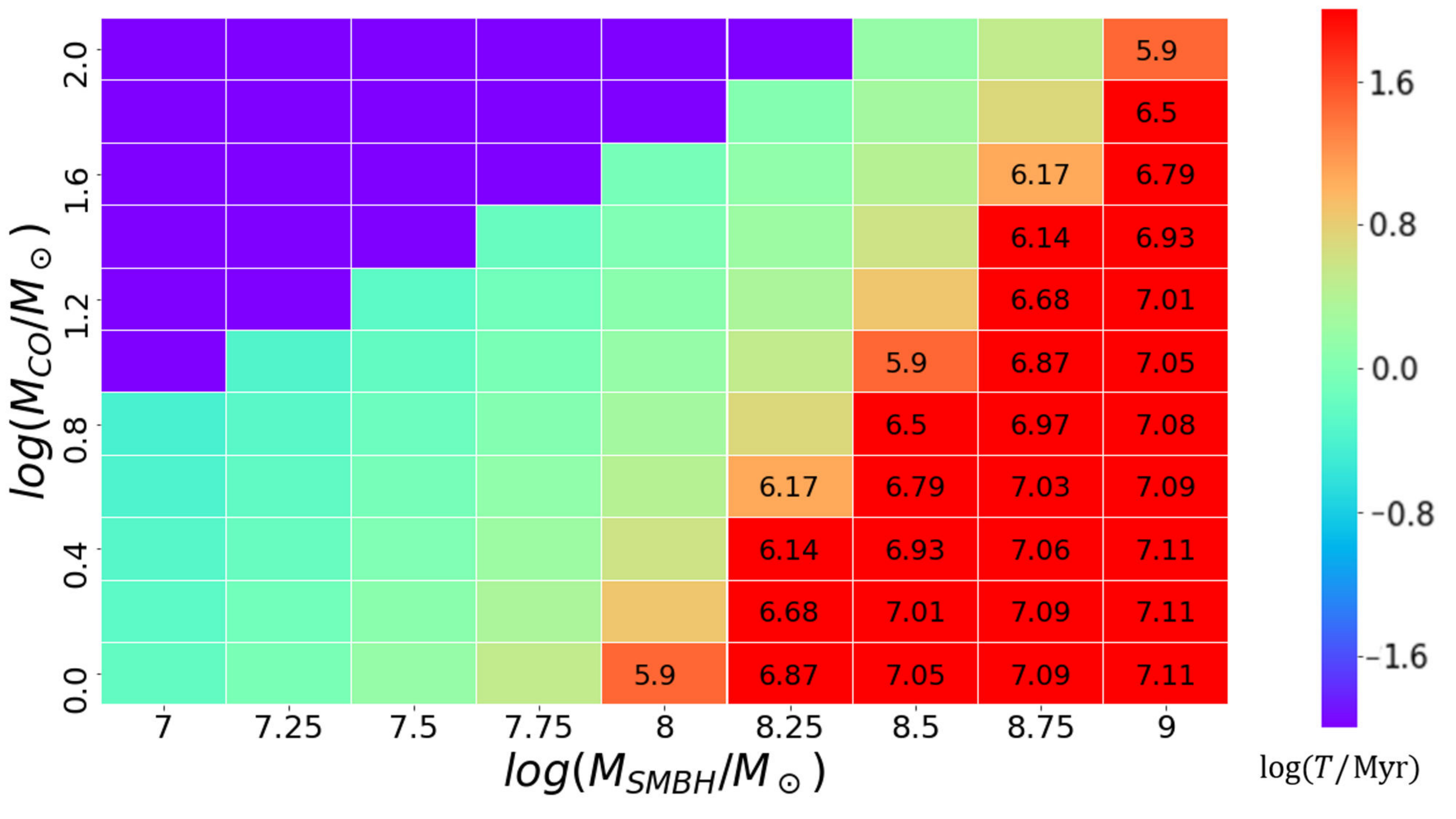}
\caption{Dependence of the migration time-scale (color coded) on the masses of
the SMBH ($M_{\rm SMBH}$) and the CO ($M_{\rm CO}$).  The other parameters are
$ \dot{M} = \dot{M}_{\rm{crit}} $ and $\alpha = 0.01$. Migration trap is
where the migration time-scale is the longest and, at the same time,
longer than the typical AGN lifetime,
$T_{\rm AGN}=10^7$ yrs, and in this case the radius of the trap is shown by the
black number (in unit of $r_s$).  The purple region shows the cases where the
gap opening criteria in Equation~(\ref{eq: crit2}) are satisfied. 
} \label{fig:result2_1} \end{figure}

To find out whether the last migration trap exits in other cases,  
we run a grid of calculations with $M_{\rm{SMBH}} = 10^7 \sim 10^9 \,
M_{\odot}$, $ \dot{M} =  0.1 \sim 10 \, \dot{M}_{\rm{crit}}$, $M_{\rm{CO}} = 1
\sim 100 \, M_{\odot}$ and $\alpha = 0.01 \sim 0.1$. We do not consider those
AGNs with higher masses and accretion rates, or the COs with lower masses,
because their are rare \citep{2009ApJ...690...20S, 2002ApJ...579..530W,
2016ARA&A..54..401O}.  Figure~\ref{fig:result2_1} shows the parameter space
where the last migration trap exists. The calculation is done assuming that $
\dot{M} = \dot{M}_{\rm{crit}} $ and $\alpha = 0.01$. The location of the
migration trap is found at the radius which satisfies two conditions: 
(i) the migration
time-scale is the longest and (ii) the migration
time-scale is also longer than $10^7$ years, the typical lifetime of AGN. The
numbers shown in the figure indicate the radii of the migration traps.
We find that larger $M_{\rm{SMBH}}$ and smaller $M_{\rm{CO}}$ are more likely to
produce migration traps.  The trapping radius is small, between
$5$ and $8$ Schwarzschild radii of the central SMBH, which is relatively
close to the ISCO ($3\,r_s$).  In the purple region (upper-left) of the figure, the
criteria shown in Equation~(\ref{eq: crit2}) are met so that a gap would form
around the CO. Type II migration would drive the COs to the ISCO and flush them
into the SMBH. The green stripe along the diagonal is the region where GW radiation is efficient and can flush the COs into the central SMBH.  

\begin{figure}
\centering
\includegraphics[width=0.5\textwidth]{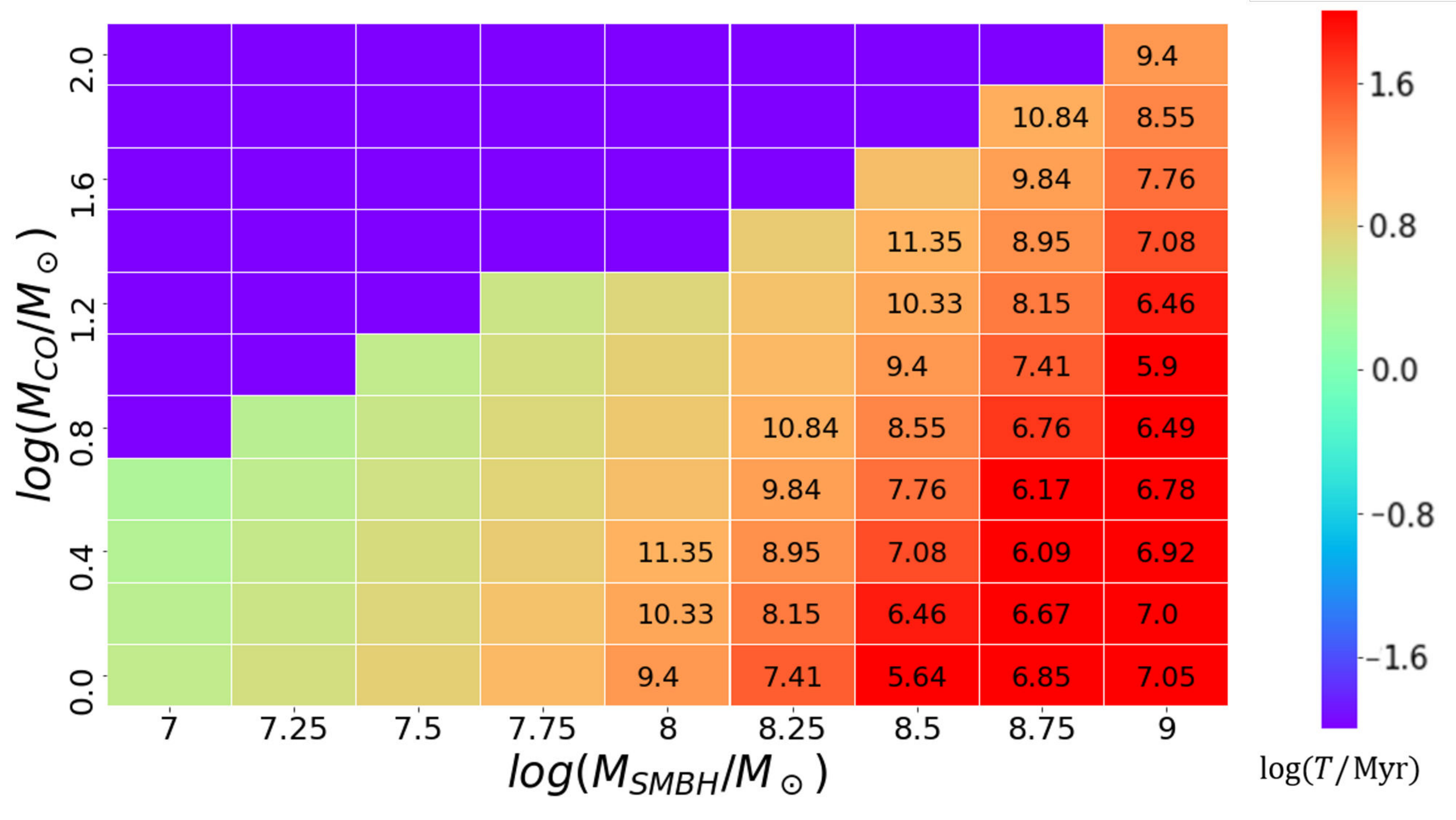}
\caption{Same as Figure \ref{fig:result2_1} but for $\alpha = 0.1$.}
\label{fig:result2_3}
\end{figure}

If we increase the viscosity parameter to $0.1$, the major change of the disc
is that the gas density drops since the advection velocity increases. 
As a result, the migration time-scales related to type I and  
headwind/tailwind both elongate. COs can now spend more time in the accretion disc,
so that a larger parameters space could host migration traps, as is shown in
Figure~\ref{fig:result2_3}.

\begin{figure*}
\begin{minipage}[t]{0.45\linewidth}
\includegraphics[width=\textwidth]{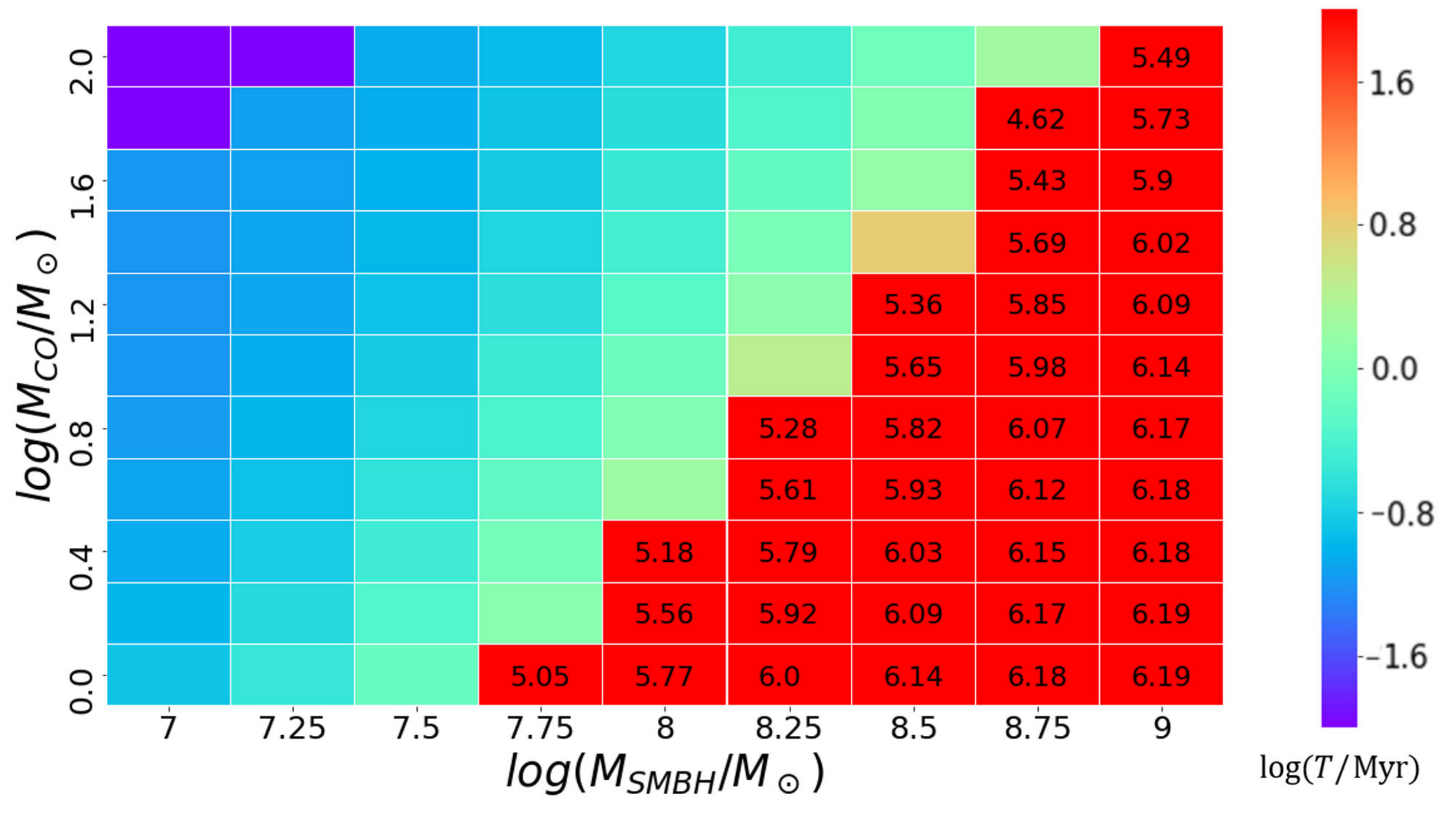}
\end{minipage}
\begin{minipage}[t]{0.45\linewidth}
\includegraphics[width=\textwidth]{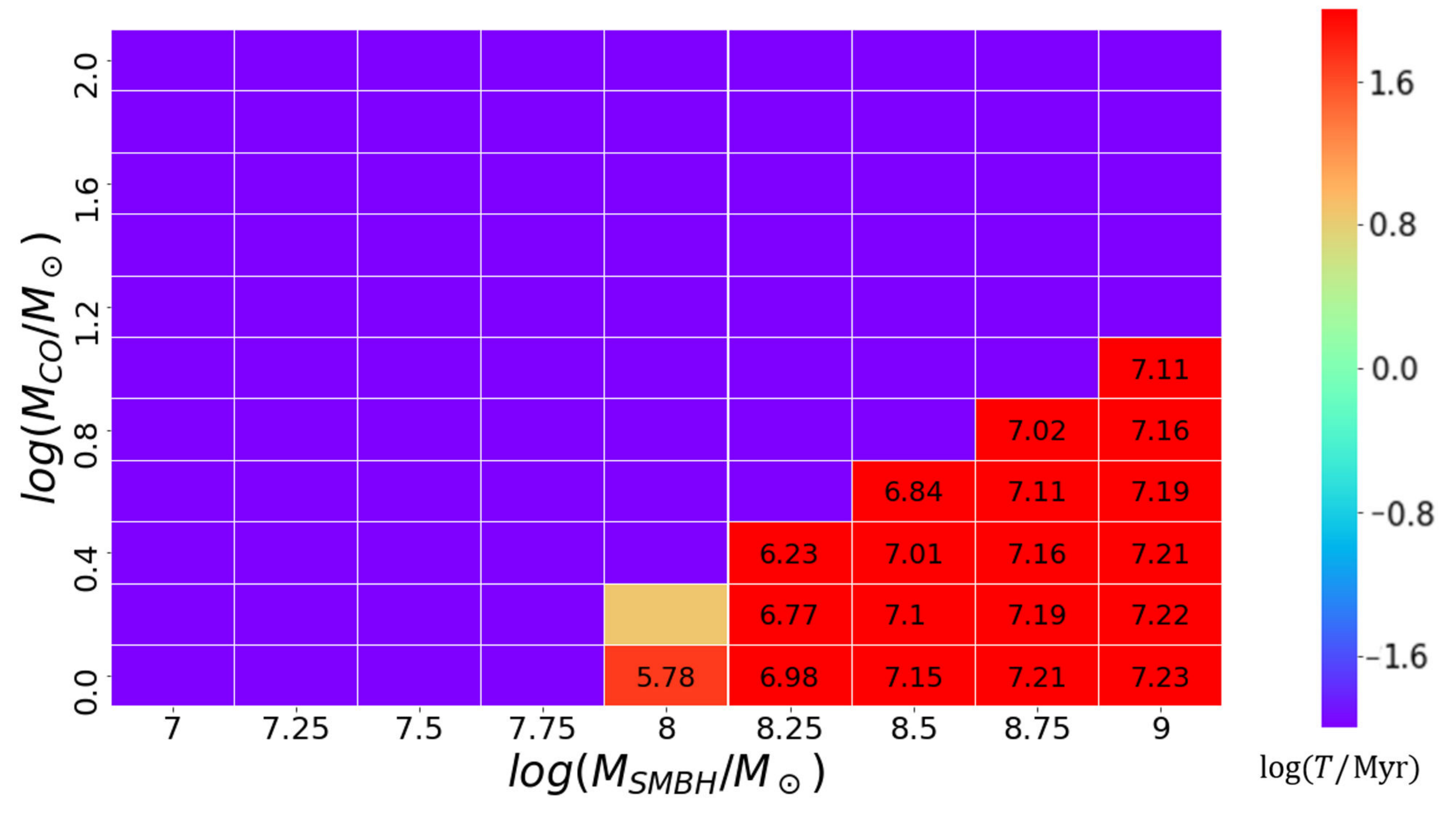}
\end{minipage}
\caption{Same as Figure \ref{fig:result2_1}, but for $ \dot{M} = 10 \dot{M}_{\rm{crit}} $ (left panel) and $ \dot{M} = 0.1 \dot{M}_{\rm{crit}} $ (right panel).}
\label{fig:result2_2}
\end{figure*}

We also studied the dependence of the location of the last migration trap on
the accretion rate of the AGN disc. The result is shown in
Figure~\ref{fig:result2_2}.  We find that higher accretion rate generally makes
the last migration trap more common in AGNs, and the location is pushed closer
to the central SMBH.  Meanwhile, lower accretion rate suppresses the occurrence
of the last migration trap because the headwind/tailwind become weaker.
These behaviors can be understood in the context of the strength of the
tailwind and its balance with the GW torque. 

\section{Discussion} \label{sec:discussion}

In this work, we have studied the migration of the COs embedded in a slim disc
around a SMBH. Such discs are expected to form in the most luminous quasars.
We paid particular attention to the non-Kepler motion of the gas in the disc
and its effect on the COs.  We find that such a disc could cause the COs to
migrate to a place close to the ISCO where the rotation velocity
of the gas transforms from sub-Keplerian to super-Keplerian, and, more importantly,
trap the COs there for as long as the lifetime of the accretion disc. We call
this place the ``last migration trap'' because inside it COs can no longer stay
for a longtime due to GW radiation.

Our results suggest that the last migration trap is more common in AGNs with
heavier SMBHs ($\ga10^8M_\odot$) and higher accretion rates
($\dot{M}\ga0.1\dot{M}_{\rm crit}$), and it traps more easily for lighter COs.
Take stellar-mass BHs with a mass of $10 M_\odot$ for example, we find that the
corresponding last migration trap exists when the SMBH is heavier than $3
\times 10^8 \, M_\odot$ and the accretion rate is higher than $1 \,
M_{\rm{crit}}$, or when the SMBH is heavier than $10^9 \, M_\odot$ and the
accretion rate is higher than $0.1 \, M_{\rm{crit}}$.  These requirements for
the central SMBH are characteristically different from those derived in the
earlier works of
\citet{1993ApJ...411..610C,1994ApJ...436..249M,1996PhRvD..53.2901C,2000astro.ph.12529C,2008AIPC.1053...33B},
where they found that the migration trap exists even when $M_{\rm
SMBH}=10^7M_\odot$ or $M_{\rm CO}=10^3M_\odot$. The discrepancy is caused by
the fact that the earlier works assumed a large angular momentum for the gas
close to the ISCO, while in our work we computed the angular momentum from a
self-consistent accretion disc model.  It is worth noting that since the
central SMBH is massive when the last migration trap exists, the orbital motion of the trapped COs
are generating GWs in a frequency band
much lower than $10^{-3}$ Hz. Such a GW signal cannot be detected by the planned
Laser Interferometer Space Antenna \citep{amaro17}.

The exact locations of the last migration traps can be found in
Figures~\ref{fig:result2_1}, \ref{fig:result2_3} and \ref{fig:result2_2}.  In
other cases, the last migration trap does not exist because the tailwind on the
CO, generated by the super-Keplerian inner disc, is too weak to counter the
inward migration caused by GW radiation. In these cases where the last
migration trap does not exit, we arrive at the same conclusion as in
\citet{2021arXiv210109146P} that the COs in AGN discs would be flushed to
the ISCO and subsequently fall into the central SMBH, though our disc model is
different from that in the previous work and we included the mini disc around
the CO in the calculation of the drag force.   

The presence of a mini disc is a key assumption in our model and hence deserves
some discussion.  Although it is well accepted that mini discs exist in those
two-body systems undergoing mass transfer, such as X-ray binaries
\citep{1976IAUS...73..237L, 1994ApJ...421..651A}, it is unclear whether the
mini disc still exists around a CO embedded in the slim disc of an AGN.  Our
assumption of the mini disc is based on the previous studies on protoplanetary
discs.  Recent numerical simulations of protoplanetary discs show that mini
discs can form around planetesimals \citep[e.g.][]{2009MNRAS.392..514M}.  In
particular, \citet{2015MNRAS.446.1026O} and \citet{2015MNRAS.447.3512O} have
shown that mini discs can form in both the wind-dominated and shear-dominated
regimes, which correspond to the scenarios depicted, respectively, by the left
and right panels of Figure~\ref{fig:differential rotation}.  The size of the
mini disc around a planetesimal is of the order of the Hills radius
\citep[e.g.][]{2015ApJ...811..101F}. However, we note that slim discs have
different geometries and thermal-dynamical properties relative to planetary
discs. Numerical simulations are needed to elucidate the structure of the gas
flow around the COs embedded in slim discs.   

In our model, although a large amount of gas will interact with the mini disc,
the mass of the disc is assumed to be constant. This assumption is based on the
observation that numerical simulations often show strong outflows from the mini
discs \citep[e.g.][]{2012ApJ...747...47T, 2015ApJ...811..101F,
2015MNRAS.446.1026O, 2015MNRAS.447.3512O}. One can also compare the internal
energy of the gas and the gravitational potential of the CO. he specific
internal energy of gas after collision is $ (c_s^2 + V_{\rm{rel}}^2)/2 $, where
the second term comes from the kinetic energy of gas before collision. The
gravitational potential energy at the edge of the mini disc is $G M_{\rm{CO}}/
R_H$. Similar to the analysis for Equation \ref{eq: cross section ratio}, 
we find that the
internal energy is larger. Since the internal energy is larger than
the potential energy, the gas tends to flow out of the potential of the CO
rather than being accreted.  Moreover, we assume that the outflow is isotropic
so that there is no net force on the mini disc as the gas flows away. The
validity of this assumption need to be checked in future numerical simulations.
Nevertheless, the details of the outflow would include an extra factor in the
expression for the drag force. This factor should be of order unity, unless the
effect of outflow somehow cancels out precisely the effect of collision which
is unlikely. 

Now we estimate the BBH merger rate contributed by the BHs in the last
migration trap. The relevant SMBHs are in the mass range of 
$10^8-10^9 M_\odot$ and have an accretion rate higher than $0.1M_{\rm crit}$.
Further taking into account the fact that LIGO/Virgo only detect
the BBHs within a redshift of $z\simeq1$, we find that the 
number density of the quasars which satisfy the above constraints is 
$n_{\rm{quasars}} = 10^{-7} {\rm{Mpc}}^{-3}$ \citep{2009ApJ...692.1388K, 2009ApJ...699..800V}. 
For each quasar, we following the result of \citep{2020ApJ...898...25T} and 
assume that a number of
\begin{equation} \label{eq: BH number}
    N_{\rm{BH}} = 3 \times 10^4 \left(\frac{M_{\rm{SMBH}}}{10^8 M_\odot}\right)^{0.827} 
\left(\frac{\dot{M}}{\dot{M}_{\rm{crit}}}\right)^{1/2}
\end{equation}
of stellar-mass BHs are captured by the accretion disc. These BHs initially form
in the nuclear star cluster, within a distance comparable to the size of the 
accretion disc from the central SMBH.    
\cite{2020ApJ...898...25T} showed that the BHs 
in the accretion disc
can effectively migrate inward and \cite{2021arXiv210109146P} further showed that they
would not be trapped by the type I migration trap. Therefore, we assume that the majority
of these BHs could reach the last migration trap and form BBHs.
With these consideration and typical parameter $M_{\rm{SMBH}} = 3 \times 10^8 M_\odot$, $\dot{M} = \dot{M}_{\rm{crit}}$, we estimate that the event rate is
\begin{equation}
    R = \frac{1}{2} n_{\rm{quasars}} N_{\rm{BH}} / T_{\rm{AGN}} \sim 0.4 \, {\rm{Gpc}}^{-3} {\rm{yr}^{-1}},
\end{equation}
where we have assumed a typical lifetime of $T_{\rm AGN}=10^7$ yrs for a quasar.  We note that this event
rate is much higher than the merger rate from another channel, i.e., tidal capture of BBHs by the SMBHs
in gas poor galactic nuclei \citep{2018CmPhy...1...53C}. Therefore, the last
migration trap is the dominant channel of producing BBH mergers in the vicinity
of SMBHs. 

The event rate is relatively low compared to the BBH merger rate inferred from
the current LIGO/Virgo events. In fact it is about $1$ per cent of the rate
inferred from LIGO/Virgo observations. Therefore, it is unlikely that all
LIGO/Virgo BBHs are produced in the last migration traps. However, as the
number of events in the O3 observing run reaches $80-90$ \citep{2020LRR....23....3A}, the chance of detecting one such event becomes considerably high.  

In fact, one LIGO/Virgo event, namely GW190521 \citep{2020PhRvL.125j1102A}, is
reported to coincident with an AGN flare and hence  may be produced inside the
accretion disc of the AGN \citep{2020PhRvL.124y1102G}, although alternative
interpretation of the AGN flare also exists \citep[e.g.][]{2020MNRAS.499L..87D}.
GW190521 is also the most massive BBH detected so far, whose BH masses prior to the
merger appear to be $66M_\odot$ and $85M_\odot$.  
Interesting, the luminosity distance of the AGN, $d_{\rm AGN}\simeq2.2$ Gpc, does not match the 
apparent distance inferred from the GW signal, $d_{\rm GW}\simeq4.5$ Gpc. For others, 
such a discrepancy would refute the association between the AGN and GW190521.
For us, however, the discrepancy can be explained using the Doppler and gravitational
redshift around the SMBH. \citet{2019MNRAS.485L.141C} showed that
the above two distances are not identical but related as $d_{\rm GW}/d_{\rm AGN}=1+z$ if Doppler and gravitational 
redshifts are important. At the ISCO, the total effect of the aforementioned 
two redshifts ($1+z$) is between $1.9$ and $3.4$, depending on the spin parameter of the SMBH. Such
a value explains well the different between $d_{\rm AGN}$ and $d_{\rm GW}$.
If GW190521 is indeed
produced in the last migration trap, the Doppler and gravitational redshifts
could increase the apparent mass by a factor of $d_{\rm GW}/d_{\rm AGN}\simeq2$ according to
the theory laid out in \citet{2019MNRAS.485L.141C}. Consequently, the real masses of
the BHs would be lowered to $(30-40) M_\odot$. Such smaller BH masses would
greatly ease the tension between GW190521 and the theoretical upper limit for
the BHs produced by stellar evolution \citep{2017ApJ...836..244W}.

The existence of the last migration trap also has important implication for the
SMBHs in normal galaxies.  For example, the center of M87 harbors a SMBH with a
mass of $6.5 \times 10^9 M_\odot$ \citep{2019ApJ...875L...1E}.  Its accretion
rate is less than $ 10^{-5} \dot{M}_{\rm{crit}}$ \citep{2021ApJ...910L..13E},
but in the past the accretion rate should be much higher, probably approaching
the Eddington limit, so that the SMBH could grow within a Hubble time to the
current mass. During such an active phase, the AGN was likely to have a last
migration trap given the large mass of the SMBH. The trap should have created a
swamp of BHs in the vicinity of the SMBH. According to Equation~(\ref{eq: BH
number}), the number of BHs can reach $4\times 10^5$ assuming
$\dot{M} = 0.1 \dot{M}_{\rm{crit}}$. Their mutual interaction could have produced
many BBHs through the mechanism presented in \citet{2019ApJ...878...85S,
2020ApJ...898...25T}.  If some BBHs survive till today, given the closeness of
M87 whose distance is about $17$ Mpc \citep{2018ApJ...856..126C}, they may be
detectable by LISA which is more sensitive to long-living BBHs than LIGO/Virgo.
 
If some BBHs in the LIGO/Virgo band are produced in the last migration
trap, it is important to find a way of identifying them. The current difficulty
lies in the shortness of the signal in the LIGO/Virgo band, which is normally
less than a second. During such a short time span, the acceleration of the
center-of-mass of the source, which is often considered as a smoking-gun
evidence of a nearby SMBH \citep[e.g.][]{bonvin17,inayoshi17,meiron16}, is too
weak to be discernible in the waveform.  However, recent theoretical works
showed that the high velocity of the BBH around the SMBH
\citep{2019PhRvD.100f3012T,2020PhRvD.101h3028T,2020arXiv201015856T} could
induce additional modes to the GW radiation.  A possible strong lensing by the
SMBH \citep{kocsis13,2020PhRvD.101h3031D} may also produce distinguishable
signatures in the waveform \citep{dai17,wang21,ezquiaga21}. Systematic search
in the LIGO/Virgo data for these predicted signatures will help elucidate the
importance of this channel of forming BBHs in the last migration trap. 

\section*{Acknowledgements}

This work is supported by NSFC grants No. 11873022 and 11991053. X.C.  is
supported partly by the Strategic Priority Research Program “Multi-wavelength
gravitational wave universe” of the Chinese Academy of Sciences (No.
XDB23040100 and XDB23010200). The computation in this work was performed on the
High Performance Computing Platform of the Centre for Life Science, Peking
University.

%%%%%%%%%%%%%%%%%%%%%%%%%%%%%%%%%%%%%%%%%%%%%%%%%%

%%%%%%%%%%%%%%%%%%%% REFERENCES %%%%%%%%%%%%%%%%%%

% The best way to enter references is to use BibTeX:

%\bibliographystyle{mnras}
%\bibliography{example} % if your bibtex file is called example.bib

% Alternatively you could enter them by hand, like this:
% This method is tedious and prone to error if you have lots of references

\bibliographystyle{mnras.bst}
\bibliography{bibbase,mybib}

%%%%%%%%%%%%%%%%%%%%%%%%%%%%%%%%%%%%%%%%%%%%%%%%%%

%%%%%%%%%%%%%%%%% APPENDICES %%%%%%%%%%%%%%%%%%%%%

%%%%%%%%%%%%%%%%%%%%%%%%%%%%%%%%%%%%%%%%%%%%%%%%%%

% Don't change these lines
\bsp	% typesetting comment
\label{lastpage}
\end{document}